\documentclass[conference]{IEEEtran}
\IEEEoverridecommandlockouts
\usepackage{cite}
\usepackage{amsmath,amssymb,amsfonts}
\usepackage{algorithmic}
\usepackage{graphicx}
\usepackage{subcaption}
\usepackage{textcomp}
\usepackage{multicol}
\usepackage[section]{placeins}
\usepackage{float}
\usepackage{stfloats}

\def\BibTeX{{\rm B\kern-.05em{\sc i\kern-.025em b}\kern-.08em
    T\kern-.1667em\lower.7ex\hbox{E}\kern-.125emX}}
\begin{document}

\title{Using Phone Sensors and an Artificial Neural Network to Detect Gait Changes During Drinking Episodes in the Natural Environment}

\author{\IEEEauthorblockN{Pedram Gharani}
\IEEEauthorblockA{\textit{School of Computing} \\
\textit{and Information}\\
\textit{University of Pittsburgh}\\
peg25@pitt.edu}
\and
\IEEEauthorblockN{Brian Suffoletto, M.D.}
\IEEEauthorblockA{\textit{Department of Emergency Medicine} \\
\textit{School of Medicine}\\
\textit{University of Pittsburgh}\\
suffbp@upmc.edu}
\and
\IEEEauthorblockN{Tammy Chung, Ph.D.}
\IEEEauthorblockA{\textit{Department of Psychiatry} \\
\textit{School of Medicine}\\
\textit{University of Pittsburgh}\\
chungta@upmc.edu}
\and
\IEEEauthorblockN{Hassan Karimi, Ph.D.}
\IEEEauthorblockA{\textit{School of Computing} \\
\textit{and Information}\\
\textit{University of Pittsburgh}\\
hkarimi@pitt.edu}

}

\maketitle
\thispagestyle{plain}
\pagestyle{plain}

\begin{abstract}
Phone sensors could be useful in assessing changes in gait that occur with alcohol consumption. This study determined (1) feasibility of collecting gait-related data during drinking occasions in the natural environment, and (2) how gait-related features measured by phone sensors relate to estimated blood alcohol concentration (eBAC). Ten young adult heavy drinkers were prompted to complete a 5-step gait task every hour from 8pm to 12am over four consecutive weekends. We collected 3-xis accelerometer, gyroscope, and magnetometer data from phone sensors, and computed 24 gait-related features using a sliding window technique. eBAC levels were calculated at each time point based on Ecological Momentary Assessment (EMA) of alcohol use. We used an artificial neural network model to analyze associations between sensor features and eBACs in training (70\% of the data) and validation and test (30\% of the data) datasets. We analyzed 128 data points where both eBAC and gait-related sensor data was captured, either when not drinking (n=60), while eBAC was ascending (n=55) or eBAC was descending (n=13). 21 data points were captured at times when the eBAC was greater than the legal limit (0.08 mg/dl). Using a Bayesian regularized neural network, gait-related phone sensor features showed a high correlation with eBAC (Pearson's r $>$ 0.9), and $>$95\% of estimated eBAC would fall between -0.012 and +0.012 of actual eBAC. It is feasible to collect gait-related data from smartphone sensors during drinking occasions in the natural environment.  Sensor-based features can be used to infer gait changes associated with elevated blood alcohol content.

\end{abstract}

\section{Introduction}
Acute alcohol intoxication is associated with numerous health risks. For example, impaired driving due to alcohol was implicated in 28\% of the 38,000 deaths from motor vehicle accidents in the US in 2016 \cite{b1}. These consequences largely stem from alcohol’s detrimental effects on psychomotor performance \cite{b2}. Compounding this risk are impaired decision-making \cite{b3} and lack of awareness of the degree of alcohol-related impairments during drinking episodes \cite{b4}.  Strategies to measure alcohol-related psychomotor impairments and provide real-time feedback to individuals could deter involvement in activities that require psychomotor function (i.e., driving), thus reducing likelihood of injury \cite{b5}.

Using alcohol consumption as a surrogate for psychomotor impairment can be prone to either underestimations (e.g. when individuals do not report alcohol consumption accurately \cite{b6}) or over-estimations (e.g. in individuals with high tolerance to the effects of alcohol \cite{b7}).  Therefore, asking individuals to input drinks in real time may not be the most accurate estimate of psychomotor impairment. One measure of psychomotor performance that is particularly sensitive to alcohol is gait. Gait requires coordination of multiple sensory and motor systems.  Both postural stability \cite{b8} and gait \cite{b9} are sensitive to blood alcohol concentration (BAC) levels.  Although law enforcement professionals have used subjective performance on a heel-to-toe tandem gait task as a field sobriety test for years, there is no current process to objectively measure aspects of gait during drinking occasions in the natural environment.

The rapid growth of smartphone ownership \cite{b10} and standard inclusion of accelerometer and gyroscope sensors within phones suggests that these devices could be useful to objectively measure gait impairment during drinking episodes. Researchers have begun to model the associations between gait abnormalities detected using smartphone sensors and either real or simulated alcohol consumption \cite{b11},\cite{b12}, but no one has determined if it is feasible to collect standardized gait data in the natural drinking environment from at-risk individuals, nor the association of gait features with BAC levels. Answers to these questions are critical if sensor based data will be used to provide feedback that is specific to drinking risk level.

	The aim of this project was to determine (1) feasibility of collecting gait-related data during a brief gait task during drinking occasions from at-risk young adults in the natural environment, and (2) how gait-related features measured through phone sensors relate to estimated blood alcohol concentration (eBAC). To accomplish these aims, we designed an iPhone app (DrinkTRAC) to collect smartphone sensor-based data on gait (3-axis accelerometer, gyroscope, magnetometer) and ecological momentary assessment (EMA) measures of self-reported number of drinks consumed each hour, from 8pm to 12am, during weekend evenings (Fridays and Saturdays). We enrolled 10 young adults with a history of heavy drinking in a repeated-measures study to provide smartphone sensor and self-report data over a period of four consecutive weeks. We used a Bayesian regularized neural network (BRNN) to perform regression analysis to examine the association of sensor data with eBACs. Results from this work could be useful in designing effective prevention interventions to reduce risky behaviors during periods of alcohol intoxication.

\section{Methods}
\subsection{Participants}
A convenience sample of young adults (aged 21--–26 years) who presented to an urban Emergency Department (ED) between February 19 and May 9, 2016 were recruited. A total of 28 medically stable ED patients who were not seeking treatment for substance use, not intoxicated, and who were going to be discharged to home, were approached by research staff. Among those eligible to be approached, 23 patients provided consent to complete an alcohol use severity screen. Those who reported recent hazardous alcohol consumption based on an Alcohol Use Disorder Identification Test for Consumption (AUDIT-C) score of $\geq$3 for women or $\geq$4 for men \cite{b13} and who drank primarily on weekends were eligible for participation. We excluded those who reported any medical condition that resulted in impaired thinking or memory or gait, those who reported past treatment for alcohol use disorder, and those without an iOS phone. A total of 10 participants met the study enrolment criteria. All participants completed informed consent protocols prior to study procedures and were provided with resources for alcohol treatment. 

\begin{table}[]
\centering
\caption{Sample descriptive statistics}
\label{table1}
\begin{tabular}{lllc}
\hline
\multicolumn{3}{l}{Characteristics} & N=10 \\ \hline
\multicolumn{3}{l}{Age in years, mean (SD)} & 23.1 (2.6) \\
\multicolumn{3}{l}{Female, n (\%)} & 7 (70\%) \\
\multicolumn{3}{l}{Race, n (\%)} &  \\
 & \multicolumn{2}{l}{African American} & 2 (20\%) \\
 & \multicolumn{2}{l}{White} & 6 (60\%) \\
 & \multicolumn{2}{l}{Other} & 2 (20\%) \\
\multicolumn{3}{l}{Hispanic Ethnicity, n (\%)} & 1 (10\%) \\
\multicolumn{3}{l}{Education, n (\%)} &  \\
 & \multicolumn{2}{l}{Some college} & 5 (50\%) \\
 & \multicolumn{2}{l}{College graduate or,post-graduate} & 5 (50\%) \\
\multicolumn{3}{l}{Employment, n (\%)} &  \\
 & \multicolumn{2}{l}{For wages} & 7 (70\%) \\
 & \multicolumn{2}{l}{Student} & 3 (30\%) \\
\multicolumn{3}{l}{Married, n (\%)} & 1 (10\%) \\
\multicolumn{3}{l}{Alcohol Use Severity (AUDIT-C score), mean (SD)} & 5 (1.3) \\
\multicolumn{3}{l}{Weight in pounds, mean (SD)} & 179 (35) \\ \hline
\end{tabular}
\end{table}

\subsection{DrinkTRAC Application}
The DrinkTRAC app was developed using Apple’s ResearchKit platform, as it allows for convenient and professional-appearing modular builds that incorporate timed psychomotor tasks. The DrinkTRAC app contained the following components: (1) timed electronic notifications, (2) a home screen, (3) a baseline questionnaire, (4) a calendar displaying completed tasks, (5) timed EMA, and (6) timed psychomotor tasks (including a 5-step gait task). The gait task was taken from the mPower GitHub site (https://github.com/ResearchKit/mPower). We altered the gait task to take less than 45 seconds to optimize completion and reduce potential for disruptions that could interfere with task performance.

\begin{figure*}[t!]

	\begin{center}
		\begin{multicols}{2}
		
    		\includegraphics[scale=.6]{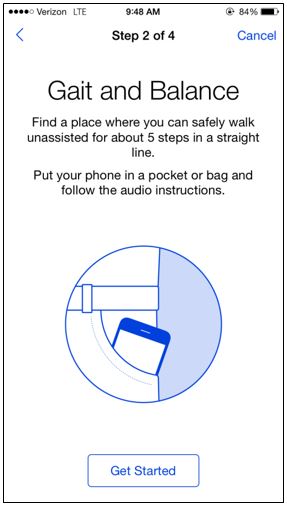}\par 
    		\subcaption{}
    		\includegraphics[scale=.6]{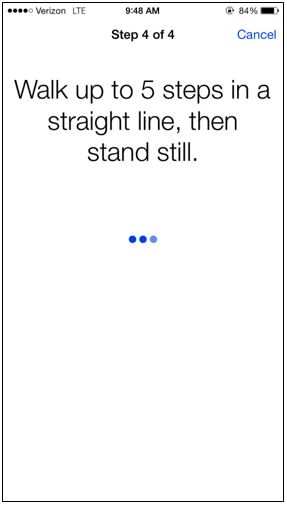}\par 
    		\subcaption{}
    	\end{multicols}
	\caption{DrinkTRAC app screenshots of the tandem gait task}
	\label{fig1}
	\end{center}

\end{figure*}

Prior to each gait task, participants were instructed to walk in a straight line for 5 steps. We advised participants not to continue if they felt that they could not safely walk 5 steps in a straight line unassisted. If participants clicked “next”, they were shown a picture of a phone and told: “Find a place where you can safely walk unassisted for about 5 steps in a straight line”, followed by the text: “Put the phone in a pocket or bag and follow the audio instructions. If you do not have somewhere to put the phone, keep it in your hand”. When the participant clicked “Get Started”, the app displayed a timer and played an audio recording of a voice counting down from 5 to 1. If the audio option was turned on, participants heard “Walk up to 5 steps in a straight line, then stand still.”  When the task was completed, participants were presented with a figure of their completion rates for the day. Figure \ref{fig1} shows DrinkTRAC app screen shots of the gait task.

\subsection{Procedure}

In the ED, participants uploaded the DrinkTRAC app to their phone. The research associate was present to ensure understanding and to observe compliance with instructions on the initial trial of the app's tasks, which were conducted in the ED. We instructed participants to refrain from any non-drinking substance use (excluding cigarette use) during the sampling days. We also informed participants that they would receive \$10 for completing the baseline survey and app-based tasks in the ED, \$10 for completing the exit survey at four weeks, and \$1 per completed EMA (up to an additional \$40).

Over four consecutive Fridays and Saturdays, every hour from 8pm to 12am, participants were sent an electronic notification to log in to the DrinkTRAC app and complete the tasks. We chose to sample data on weekend evenings, given that this is a time when young adults typically drink alcohol \cite{b14}. We collected data hourly from 8pm to 12am on those nights, with an intention of capturing both the ascending (when eBAC is rising) and descending (when eBAC is decreasing) limbs of alcohol intoxication. We used fixed hourly assessment times, given that they would provide a predictable framework for participants and would allow us to more easily calculate eBAC changes over the course of the evening.

\subsection{Data Processing}
\subsubsection{Estimated Blood Alcohol Concentration}
We calculated eBAC during each hour when data was available using a formula created by Matthews and Miller \cite{b15}. This formula was found to have a significantly stronger intraclass correlation with breath alcohol concentrations (criterion standard) than did other equations when measured after an uncontrolled episode of drinking \cite{b16}.

\subsubsection{Sensor Data Feature Extraction}
We collected 3-axis acceleration and angular velocity data, sampled at a frequency of 100 Hz for 30 seconds during the tandem-gait task.  We extracted features from sensors that can describe the properties of gait. We consider four features, i.e., mean, standard deviation, correlation, and energy. Energy is a feature extracted from a signal through a Fast Fourier Transformation (FFT) from time domain into frequency domain. Energy feature can reveal how energy in the signal is distributed over its various frequency components. Energy is the sum of the squared discrete FFT coefficient magnitudes of the signal. The sum was divided by the length of the window for normalization. In a formal notation, if  $x_1,x_2, \cdots$ are the FFT components of the window then $Energy= \frac{\sum\limits_{i=1}^{\left| w \right|}{\left|x_i \right|}^2}{\left| w \right|}$. Energy value differs for each activity; therefore, it can be used to determine discrepancies among certain movement patterns \cite{b17}.  In order to study the properties of each signal, we broke up time into smaller segments using a sliding window, where in each new "window" has 50\% overlap with the previous one, and then calculations are performed on the "windowed" signal. All measurements are in three dimensions; thus, resulting in a total of 24 possible features. The efficiency of these features for characterizing gait has been discussed in (Bao and Intille 2004 \cite{b18}, Ravi et al. 2005 \cite{b19}). 

We removed the effects of gravity in order to measure acceleration of the device (linear acceleration). An accelerometer is subject to dynamic (or external) and static (gravity) accelerations. Therefore, in order to measure the real acceleration of the device, the contribution of the static acceleration which is the effect of Earth's gravity on the device must be removed. Hence, for the purpose of obtaining linear acceleration, other sensors such as gyroscope and magnetometer were utilized to derive accurate values of linear acceleration by applying a high-pass filter. We also measured the attitude of the device, which is the computed device orientation using the accelerometer, magnetometer, and gyroscope. These values yield the Euler angles of the device.

\subsection{Bayesian Regularized Neural Network (BRNN) for eBAC Regression}

We applied a neural network model to estimate the association between sensor-based gait features and eBAC. One of the advantages of using neural networks for regression and predicting values is that it uses a nonlinear sigmoid function in a hidden layer, which enhances its computational flexibility, as compared to a standard linear regression model \cite{b20}. We first used multilayer perceptron (MLP), a BRNN, to model the nonlinear relationships between the extracted gait features and the output (eBAC value), with nonlinear transfer functions. A schematic diagram of our MLP structure can be seen in Appendix. We then tested the BRNN on a training set, and computed the corresponding parameters of the network, such as weights and learning rate. We used the Levenberg-Marquardt optimization algorithm to find the minimum of the multivariate function \cite{b21}.  We compared the performance of MLP with support vector machine (SVM) and linear regression by examining correlation coefficient, mean absolute error, root mean squared error, relative absolute error, and root relative squared error. We then examined the correlation between MLP model-predicted eBAC values based on sensor features and actual eBAC values. We examined a histogram of errors between predicted eBAC and actual eBAC to determine frequency of outliers and potential misclassification of gait impairment relation to eBAC.

\begin{figure}[h!]
\begin{center}
\includegraphics[width=\linewidth]{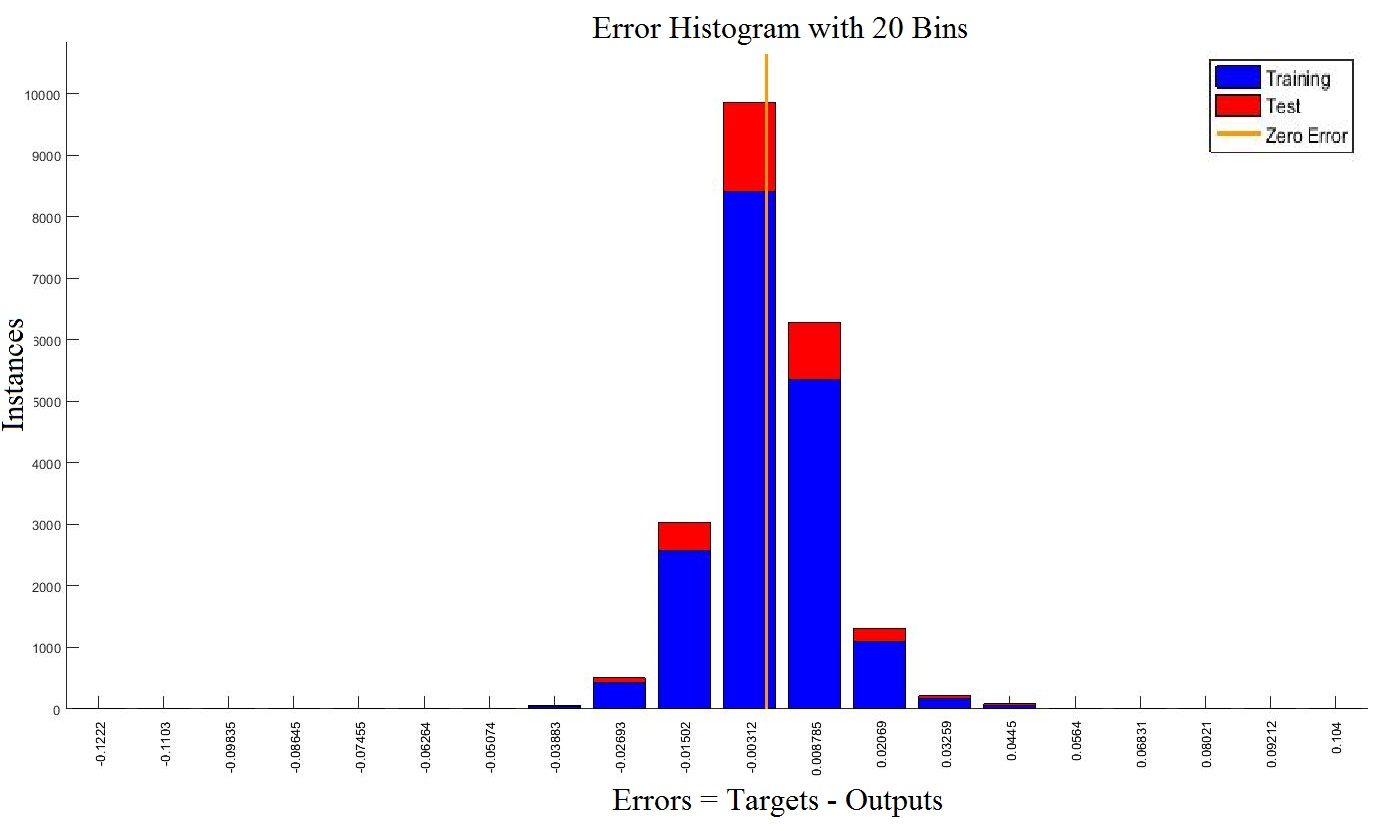}
\end{center}

\caption{Error histogram visualized errors between target values and predicted values after training a feedforward neural network with 20 bins}

\label{fig2}
\end{figure}
\section{Results}

Ten participants provided a total of 128 unique data points. Table \ref{table1} shows the baseline characteristics of the cohort.  We captured 38 unique drinking episodes, with each participant reporting at least 3 drinking episodes. Almost half of the gait tasks (n=60, 46.9\%) were completed either prior to drinking or on non-drinking evenings, 55 (43.0\%) were completed on the ascending eBAC limb, and 13 (10.1\%) were completed on the descending eBAC limb.  On a drinking day, participants reported consuming a mean of 3.6 (SD=2.2; range: 1-10 drinks). The mean eBAC was 0.04 (SD=0.05), with a peak of 0.23. 21 (15.2\%) data points were recorded at times when the eBAC was greater than the legal limit (0.08 mg/dl).

\subsection{Regression Results}
Based on comparison results of correlation coefficient, mean absolute error, root mean squared error, relative absolute error, and root relative squared error, MLP outperformed SVM and linear regression models, as shown in Table \ref{table2}. The correlation between MLP model-predicted eBAC values based on gait sensor features and actual eBAC values for training (a), testing (b), and all data (c) are also shown in this table. In the testing dataset, we found moderate variability in predicted-actual eBAC values that did not differ by actual eBAC.

\begin{table*}[b!]
\centering
\caption{Comparison of different regression techniques for BAC estimation}
\label{table2}
\begin{tabular}{cccccc}
\hline
Regression,technique & Correlation coefficient & Mean absolute error & Root mean squared error & Relative absolute error & Root relative squared error \\ \hline
MLP & 0.9009 & 0.0174 & 0.0226 & 40.6458 \% & 43.8853 \% \\
SVM & 0.3939 & 0.0362 & 0.0482 & 84.5348 \% & 93.6504 \% \\
Linear,Regression & 0.4367 & 0.0378 & 0.0463 & 88.2747 \% & 89.9583 \% \\ \hline
\end{tabular}
\end{table*}

\begin{figure*}[t!]

	\begin{center}
		\begin{multicols}{3}
		
    		\includegraphics[width=\linewidth]{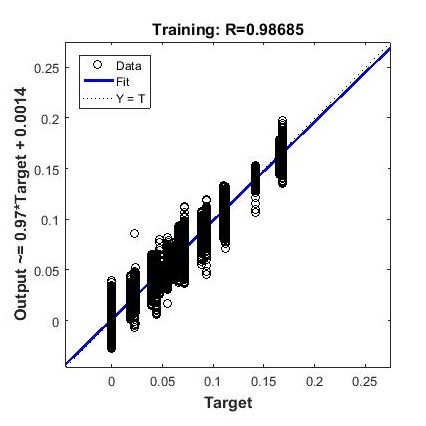}\par 
    		\subcaption{}
    		\includegraphics[width=\linewidth]{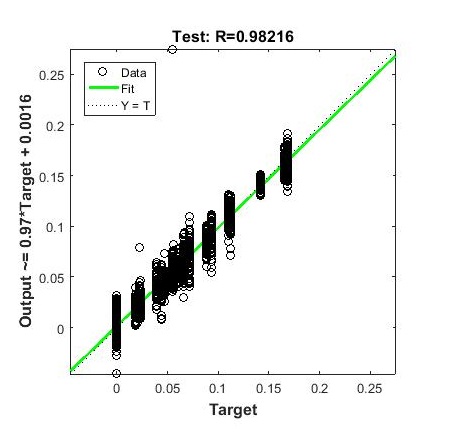}\par 
    		\subcaption{}
    		\includegraphics[width=\linewidth]{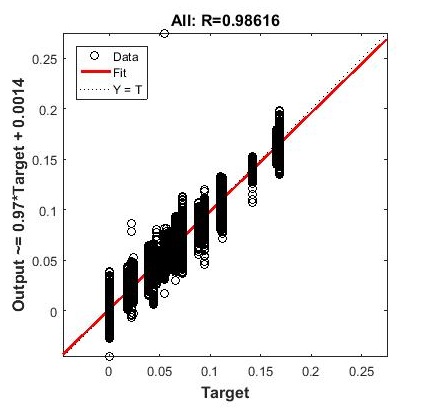}\par 
    		\subcaption{}
    	\end{multicols}
	\caption{The three plots represent the training, test, and all data}
	\label{figRegRes}
	\end{center}

\end{figure*}

The histogram of errors is shown in Figure \ref{fig2} where the blue bars represent training data and the red bars represent testing data. The histogram can give an indication of outliers, which are data points where the fit is significantly worse than that of most of the data. In this case, we can see that while $>$95\% of errors between estimated eBAC and actual eBAC fall between -0.012 and +0.012, there are some training points and just a few test points that are outside of that range. These outliers are also visible on the testing regression plot (Figure  \ref{figRegRes}). If the outliers are valid data points but are unlike the rest of the data, then the network is extrapolating for these points. This means that more data similar to the outlier points should be considered in training the model. Very rarely ($<$20\%) would an individual who had an actual eBAC value greater than the legal limit ($\geq$0.08 mg/dl) be classified as having a eBAC less than the legal limit (eBAC$<$0.08 mg/dl).

\section{Discussion}
In this pilot study, we found that it is feasible to collect gait-related data from smartphone sensors during drinking occasions in the natural environment in a cohort of young adult drinkers.  Each participant completed the 5-step gait task on at least 3 different weekend evenings, but not surprisingly, there were many missing data points later in evenings. We also found that neural network algorithms using smartphone accelerometers and gyroscope data can produce accurate prediction models to infer elevated blood alcohol concentration. Based on our final model, although there was statistical variability in predicted eBAC compared to actual eBAC values, there appears to be rare occurrence of mis-classification that would be considered clinically significant.  These results contribute to the eventual goal of being able to provide real-time feedback to individuals during drinking occasions on gait impairment, which could deter involvement in activities like driving, thus reducing the likelihood of alcohol-related injury.
Our findings concur with early pilot work by Kao et al. \cite{b22} who showed good agreement between gait captured by inertial sensors and alcohol consumption (coded Yes or No) and Arnold et al. \cite{b11} who used machine learning algorithms of smartphone inertial sensors to determine the number of drinks (not BAC) in a few healthy volunteers. It also expands upon later research \cite{b12} showing that more advanced sensor-based measures including gait sway can improve algorithm performance when tested against simulated intoxication using visual-altering goggles.

Our approach differs from prior work in several important ways. First, we studied a relevant target population (young adult heavy drinkers) who would benefit most from a prevention-oriented intervention, as opposed to healthy volunteers. Second, we studied feasibility of task completion in real drinking environments.  Third, we used a widely available software platform for collecting movement data (Apple ResearchKit), as opposed to an external sensor device or a proprietary app, allowing for replication. Fourth, we calculated eBAC using established formulas, thus providing a more accurate representation of actual blood alcohol concentration than drink counting alone.  Fifth, we standardized the gait task, thus removing random variability in naturalistic walking. Sixth, we used not only accelerometers to understand movement data, but also considered gyroscope and magnetometer measurements.  Finally, we used a sliding window technique for extracting features and feeding MLP, which outperformed the other evaluated approaches. 

This pilot study is limited by the small sample size and by the amount of missing gait task data, particularly for the descending limb of alcohol intoxication. We paid participants to complete tasks, which likely artificially inflated completion rates, suggesting that completion rates would be even lower under more naturalistic conditions. In exit interviews, we found that one major barrier to task completion was the requirement to enter a password every time a task was to be initiated. We understand that it is not practical to expect individuals to complete a walking task while they are socially engaged. Future studies could sample movement-based sensor data during natural gait and examine patterns associated with alcohol consumption. Also, findings may not be applicable to other populations, such as young adults with less severe alcohol use or other age groups such as adolescents. The majority of participants were female and white, limiting generalizability. The DrinkTRAC app was made only for iOS devices, which affected study eligibility, and limits generalizability of results to other mobile devices. Self-report of alcohol use using EMA has demonstrated reliability and validity \cite{b23}, but may be subject to bias. Future work could use transdermal alcohol sensors to validate findings and EMA schedule flexibility to reduce missing data.

\section{Conclusion and Future Direction}

This work provides initial support for the utility of using movement analysis in the real world to infer elevated blood alcohol concentration. A smartphone application (DrinkTRAC) that deployed a brief tandem gait task was completed by young adults in real-world drinking occasions. Using a Bayesian regularized neural network, we modeled and fit a curve using phone sensor data to accurately predict blood alcohol concentration during drinking episodes in the natural environment.


\begin{thebibliography}{00}

\bibitem{b1}National Center for Statistics and Analysis. (2017, October). 2016 fatal motor vehicle crashes: Overview. (Traffic Safety Facts Research Note. Report No. DOT HS 812 456). Washington, DC: National Highway Traffic Safety Administration.

\bibitem{b2}Christoforou, Z., Karlaftis, M.G., and Yannis, G., 2013. Reaction times of young alcohol-impaired drivers. Accident Analysis \& Prevention, 61, 54–62.
\bibitem{b3}Steele, C.M. and Josephs, R.A., 1990. Alcohol myopia: Its prized and dangerous effects. American Psychologist, 45 (8), 921.
\bibitem{b4}Morris, D.H., Treloar, H.R., Niculete, M.E., and McCarthy, D.M., 2014. Perceived danger while intoxicated uniquely contributes to driving after drinking. Alcoholism: clinical and experimental research, 38 (2), 521–528.
\bibitem{b5}Shults, R.A., Elder, R.W., Sleet, D.A., Nichols, J.L., Alao, M.O., Carande-Kulis, V.G., Zaza, S., Sosin, D.M., Thompson, R.S., on Community Preventive Services, T.F., and others, 2001. Reviews of evidence regarding interventions to reduce alcohol-impaired driving. American journal of preventive medicine, 21 (4), 66–88.
\bibitem{b6}Gual A, Ángel Arbesú J, Zarco J, Balcells-Oliveró MLM, López-Pelayo H, Miquel L, Bobes J Risky Drinkers Underestimate their Own Alcohol Consumption. Alcohol Alcohol. 2017 Jul 1;52(4):516-517l.
\bibitem{b7}Brumback T, Cao D, King A. Effects of alcohol on psychomotor performance and perceived impairment in heavy binge social drinkers. Drug Alcohol Depend. 2007 Nov 2;91(1):10-7.
\bibitem{b8}Nieschalk, M., Ortmann, C., West, A., Schmäl, F., Stoll, W., and Fechner, G., 1999. Effects of alcohol on body-sway patterns in human subjects. International journal of legal medicine, 112 (4), 253–260.
\bibitem{b9}Jansen, E.C., Thyssen, H.H., and Brynskov, J., 1985. Gait analysis after intake of increasing amounts of alcohol. International Journal of Legal Medicine, 94 (2), 103–107.
\bibitem{b10}Anderson, M. and Rainie, L., 2015. Pew Research Center. Technology Device Ownership: 2015, 29.
\bibitem{b11}Arnold, Z., LaRose, D., and Agu, E., 2015. Smartphone Inference of Alcohol Consumption Levels from Gait. Proceedings - 2015 IEEE International Conference on Healthcare Informatics, ICHI 2015, 417–426.
\bibitem{b12}Aiello and Agu, 2016. Investigating postural sway features, normalization and personalization in detecting blood alcohol levels of smartphone users. In: 2016 IEEE Wireless Health (WH). IEEE, 1–8.
\bibitem{b13}Bradley KA, DeBenedetti AF, Volk RJ, Williams EC, Frank D, Kivlahan DR.  AUDIT-C as a brief screen for alcohol misuse in primary care. Alcohol Clin Exp Res. 2007; 31:1208–1217.
\bibitem{b14}Del Boca, F.K., Darkes, J., Greenbaum, P.E., and Goldman, M.S., 2004. Up close and personal: Temporal variability in the drinking of individual college students during their first year. Journal of consulting and clinical psychology, 72 (2), 155–164.
\bibitem{b15}Matthews, D.B. and Miller, W.R., 1979. Estimating blood alcohol concentration: Two computer programs and their applications in therapy and research. Addictive behaviors, 4 (1), 55–60.
\bibitem{b16}Hustad, J.T.P. and Carey, K.B., 2005. Using calculations to estimate blood alcohol concentrations for naturally occurring drinking episodes: a validity study. Journal of Studies on Alcohol, 66 (1), 130–138.
\bibitem{b17}Gharani, P. and Karimi, H., 2017.  Context-aware obstacle detection for navigation by visually impaired. Image and Vision Computing, 64, 103-115.
\bibitem{b18}Bao, L. and Intille, S.S., 2004. Activity recognition from user-annotated acceleration data. In: Pervasive computing. Springer, 1–17.
\bibitem{b19}Ravi, N., Dandekar, N., Mysore, P., and Littman, M.L., 2005. Activity recognition from accelerometer data. In: AAAI. 1541–1546.
\bibitem{b20}Burden, F. and Winkler, D., 2008. Bayesian Regularization of Neural Networks. 23–42.
\bibitem{b21}Lourakis, M.I., 2005. A Brief Description of the Levenberg-Marquardt Algorithm Implemened by levmar. Matrix, 3, 2.
\bibitem{b22}Kao, H.L., Ho, B.J., Lin, A.C., and Chu, H.H., 2012. Phone-based gait analysis to detect alcohol usage. Proceedings of the 2012 ACM Conference on Ubiquitous Computing - UbiComp ’12, 661.
\bibitem{b23}Shiffman S. Ecological momentary assessment (EMA) in studies of substance use. Psychological assessment. 2009;21(4):486-497.


\end{thebibliography}
\end{document}